\newcommand{\Msolar}{M$_{\odot}$}

\documentclass{iaus}
\usepackage{natbib}
\usepackage{graphicx}
\usepackage{amsmath}

\title[IAUS 266.~~$N$-body Model of NGC 188] 
{The Progeny of Stellar Dynamics and Stellar Evolution within an $N$-body Model of \\ NGC 188}

\author[Geller, Hurley \& Mathieu]   
{Aaron M.~Geller$^1$, Jarrod R.~Hurley$^2$
 \and Robert D.~Mathieu$^1$}

\affiliation{$^1$University of Wisconsin - Madison, 475 N. Charter St., Madison WI 53706, USA 
\\[\affilskip]
$^2$Swinburne University of Technology, PO Box 218, Hawthorn, VIC 3122, Australia}

\pubyear{2009}
\volume{266}  
\pagerange{1--6}
\jname{Star Clusters: Basic Galactic Building Blocks}
\editors{R. de Grijs \& J.~R.~D. L\'epine, eds.}
\begin{document}

\maketitle

\begin{abstract}

We present a direct $N$-body simulation modeling the evolution of the old (7 Gyr) open cluster
NGC 188.  This is the first $N$-body open cluster simulation whose initial binary population is 
directly defined by observations of a specific open cluster: M35 (150 Myr).  We compare
the simulated color-magnitude diagram at 7 Gyr to that of NGC 188, and discuss the blue stragglers
produced in the simulation.  We compare the solar-type main sequence binary period and eccentricity 
distributions of the simulation to detailed observations of similar binaries in NGC 188.  
These results demonstrate the importance of detailed observations in guiding $N$-body open cluster
simulations.  Finally, we discuss the implications of a few
discrepancies between the NGC 188 model and observations and suggest a few
methods for bringing $N$-body open cluster simulations into better agreement with observations.

\end{abstract}

\firstsection 
\section{Introduction}
\label{intro}

Recently, sophisticated $N$-body simulations have incorporated stellar dynamics and stellar evolution 
self consistently, gaining the ability to accurately model open cluster sized populations 
($N \sim 10^4$) of single and binary stars through many billions of years.
(e.g. \texttt{NBODY6}, \citealt{aarseth:03}, with stellar and binary 
evolution included by \citealt{hurley:00,hurley:02}).  Concurrently, large observational surveys
of open clusters, like the WIYN Open Cluster Study \citep[WOCS;][]{mathieu:00}, are coming to maturity,
providing comprehensive databases of open cluster characteristics, including detailed information 
on their binary populations.  These observations, and specifically those of the binaries, provide
important tests and guidance for $N$-body simulations.  
In this paper, we describe our first step in utilizing this wealth of data
to help guide an $N$-body simulation with the aim of creating a realistic model of NGC 188.

\section{$N$-body Model of NGC 188}
\label{model}

\begin{figure}[!ht]
\begin{center}
\begin{tabular}{ll}
\includegraphics[width=0.5\linewidth]{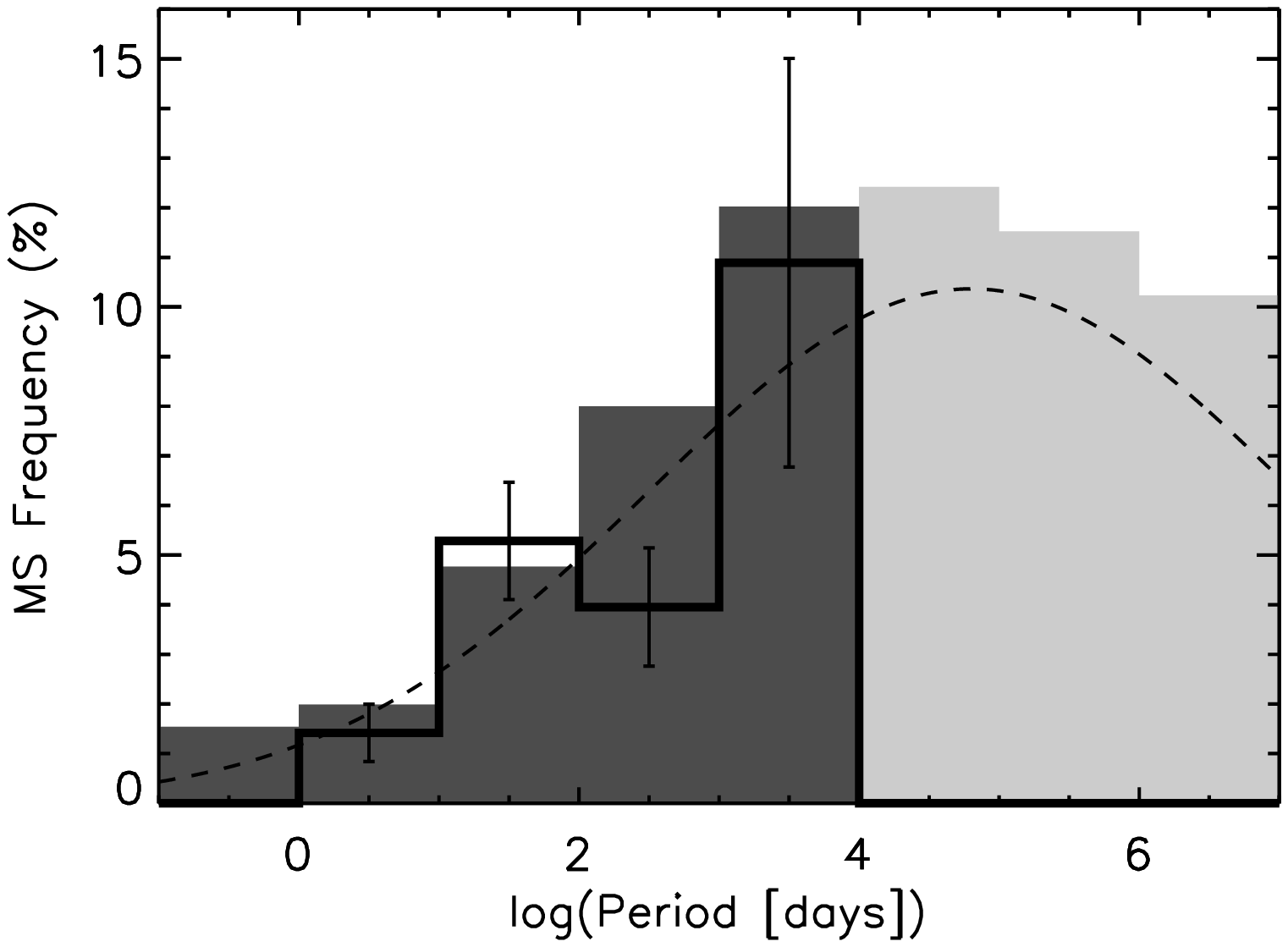} & \includegraphics[width=0.5\linewidth]{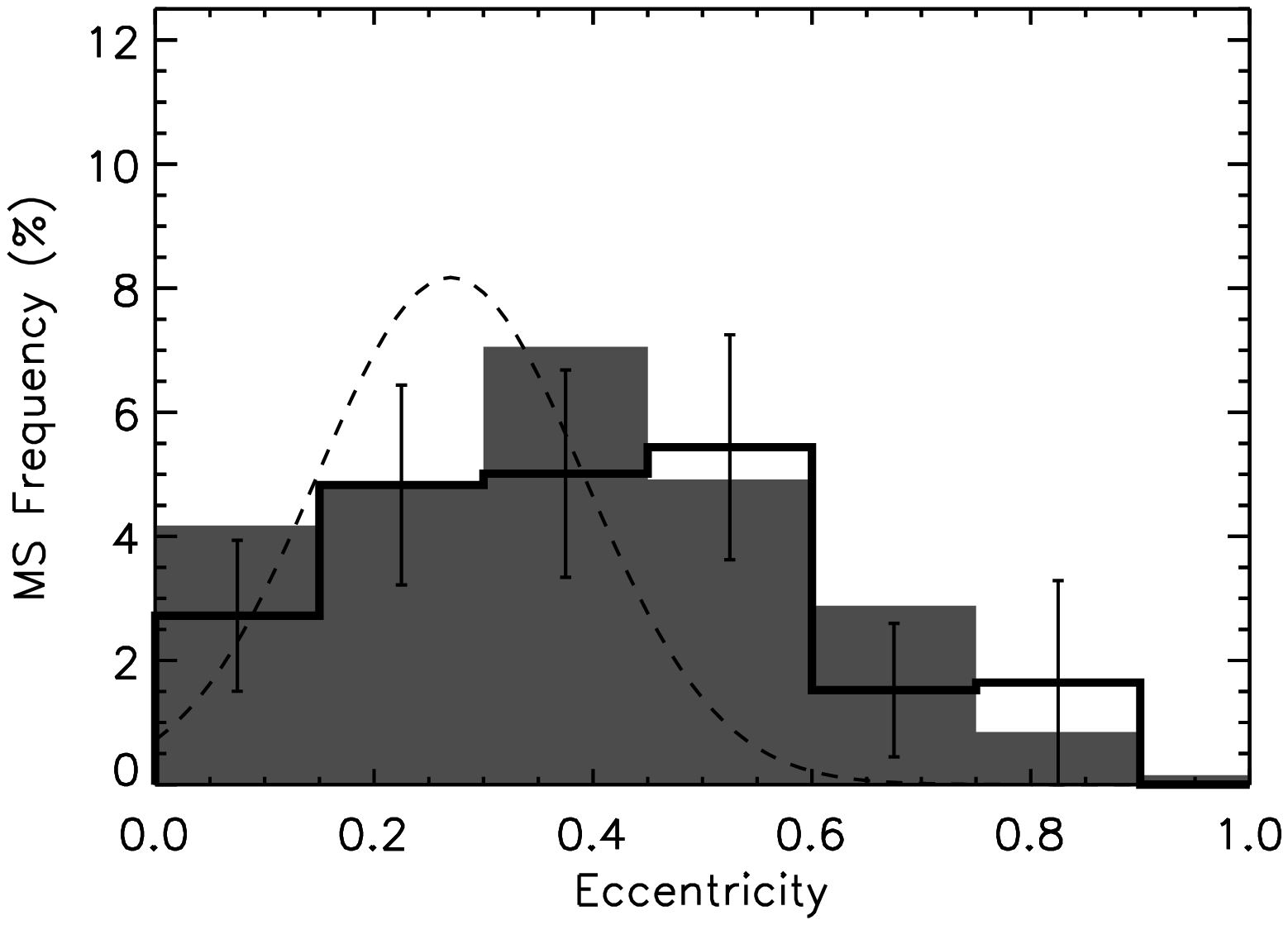} \\
\end{tabular}
\caption{\label{M35fig}
Period (\textit{left}) and eccentricity (\textit{right}) binary frequency distributions comparing the primordial 
binary population in the simulation (filled histograms) to the observed MS binary population of M35 (150 Myr; 
solid black lines).  
The M35 observed distributions have been corrected for incompleteness to show all binaries with 
$P < 10^4$ days (excluding those with periods below the circularization period in the eccentricity distribution 
plot).  The primordial binary distributions used in the simulation are plotted with dark gray for
observationally detectable binaries ($P < 10^4 $ days), and light gray for longer-period binaries. 
We also show the \citet{duquennoy:91} Galactic field binary frequency distributions in the dashed lines for 
comparison.}
\end{center}
\end{figure}

The groundbreaking \citet{hurley:05} model of M67 is a clear example of the potential for 
$N$-body simulations to model real open clusters, and thereby address long-standing questions 
about the evolution of a binary population, origins of blue stragglers (BSs), and specific observed 
characteristics of a given open cluster.  Hurley et al.~had very limited knowledge of the 
real binary population in M67, and thus aimed to match other important observables like the 
number of BSs, the spatial profile, etc. Today, through the WOCS survey,
the work at the Harvard-Smithsonian Center for Astrophysics \citep[e.g.][]{latham:06} and others, 
we now know a great deal 
about the M67 binaries.  It is now clear that, though \citet{hurley:05} were successful in matching
many observed characteristics of M67, the binary frequency at 4 Gyr was too high and the simulation
contained too many short-period binaries throughout the evolution of the cluster \citep[see][]{geller:09b}. 
The binary population has a significant
effect on the formation mechanisms and rates for BSs as well as the dynamical evolution of the
entire cluster.  Therefore, correctly modeling the binary population is one of the most important 
ingredients in creating a realistic open cluster model.  

The old (7 Gyr) open cluster NGC 188, like M67 (4 Gyr), has a rich and well-studied binary population 
\citep[e.g.,][]{geller:09a,geller:09b}, a large population of BSs \citep{mathieu:09}, and a wealth 
of observations from the WOCS group and others \citep[see][]{geller:08}.
Indeed, NGC 188 is quickly becoming one of the most well studied and important open clusters for understanding 
the origins of BSs, the characteristics of an evolved binary population and the dynamical 
state of an evolved open cluster.  These data provide ideal and much-needed guidance 
for an $N$-body simulation of NGC 188.  

\subsection{Simulation Method}
\label{method}

All simulations were run using the \texttt{NBODY6} code \citep{aarseth:00}, with slight modifications added by 
the authors to define the initial binary population and output format.  All stellar evolution is incorporated
as in \citet{hurley:00} and binary evolution, including tides, mass transfer, etc., is treated as in 
\citet{hurley:02}. BSs are modeled in the same manner as in \citet{hurley:05}.  
The simulations were executed on the supercomputer at the Swinburne Centre for Astrophysics 
and Supercomputing.  

The NGC 188 simulation was initialized with a mass of 23,870 \Msolar~($N$ = 24,000 stars) chosen
from an IMF of \citet{kroupa:01} and distributed according to a Plummer profile. 
We assume a solar metallicity for the stars in the model \citep{sarajedini:99}, and place the cluster
in orbit around a point-mass Galactic potential according to the observed orbit of NGC 188 \citep{carraro:94}.   
In the following, we discuss the binary population of the model in some detail, 
focusing on the period and eccentricity distributions and leaving more thorough discussions of methods and results
for future papers.

\begin{figure}[!ht]
\begin{center}
\includegraphics[width=1.0\linewidth]{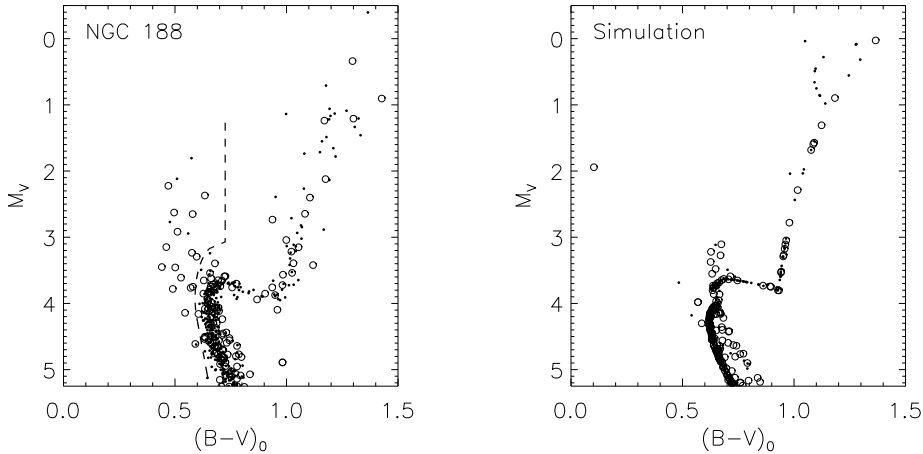}
\caption{\label{NGC188fig1}
Color-magnitude diagrams of NGC 188 (\textit{left}) and the NGC 188 simulation at 7 Gyr (\textit{right}).  The single stars are plotted in the
small black points, and the detected binaries are plotted
with the larger open circles.
The NGC 188 BSs are identified as being to the left of the dashed 
line (as in \citealt{geller:08}), and within the simulation BSs are
identified as being MS stars with masses greater than the turnoff mass (and 
reside in a similar region to observed BSs in NGC 188).  
The simulation succeeds in reproducing the overall form of the observed single and 
binary star sequence, but has not created the large number of observed BSs.}
\end{center}
\end{figure}

\subsection{Primordial Binary Population and Observations of M35}
\label{comp35}

Choosing the initial parameters defining a binary population has long been derived from theoretical
analyses and predictions, as we have lacked sufficiently detailed observations of young open cluster binary 
populations to help guide these choices.  
The WOCS survey of M35 \citep{geller:09c} now allows for a major step forward; we can use these 
binaries as a proxy for the primordial binary population in this simulation.
There is no \textit{a priori} reason to presume that the M35 binaries will 
evolve into the NGC 188 binaries.  Still, the detailed observations of the M35 binaries
allow us to create the first $N$-body model whose initial binary population
is \textit{directly defined by observations} and to study the evolution of these binaries for the entire
age of NGC 188 (7 Gyr).



M35 is observed to have a main-sequence (MS) binary frequency of 22 $\pm$ 3 \% out to periods of 10$^4$ days \citep{geller:09c}; 
we have chosen to initialize the simulation with a binary frequency of 25\% out to the same period range.
In Figure~\ref{M35fig} we show the M35 incompleteness corrected 
binary period and eccentricity distributions (black line) as 
compared to the initial conditions used in the NGC 188 model (filled histograms).  As we only 
detect binaries with periods $< 10^4$ days, we utilize the Galactic field binaries
\citep{duquennoy:91} to extend the period distribution to longer periods.
Thus we input a primordial binary 
population with a log-normal period distribution and a Gaussian eccentricity distribution. 

\subsection{Comparing to Observations of NGC 188}
\label{comp188}

\begin{figure}[!ht]
\begin{center}
\begin{tabular}{ll}
\includegraphics[width=0.5\linewidth]{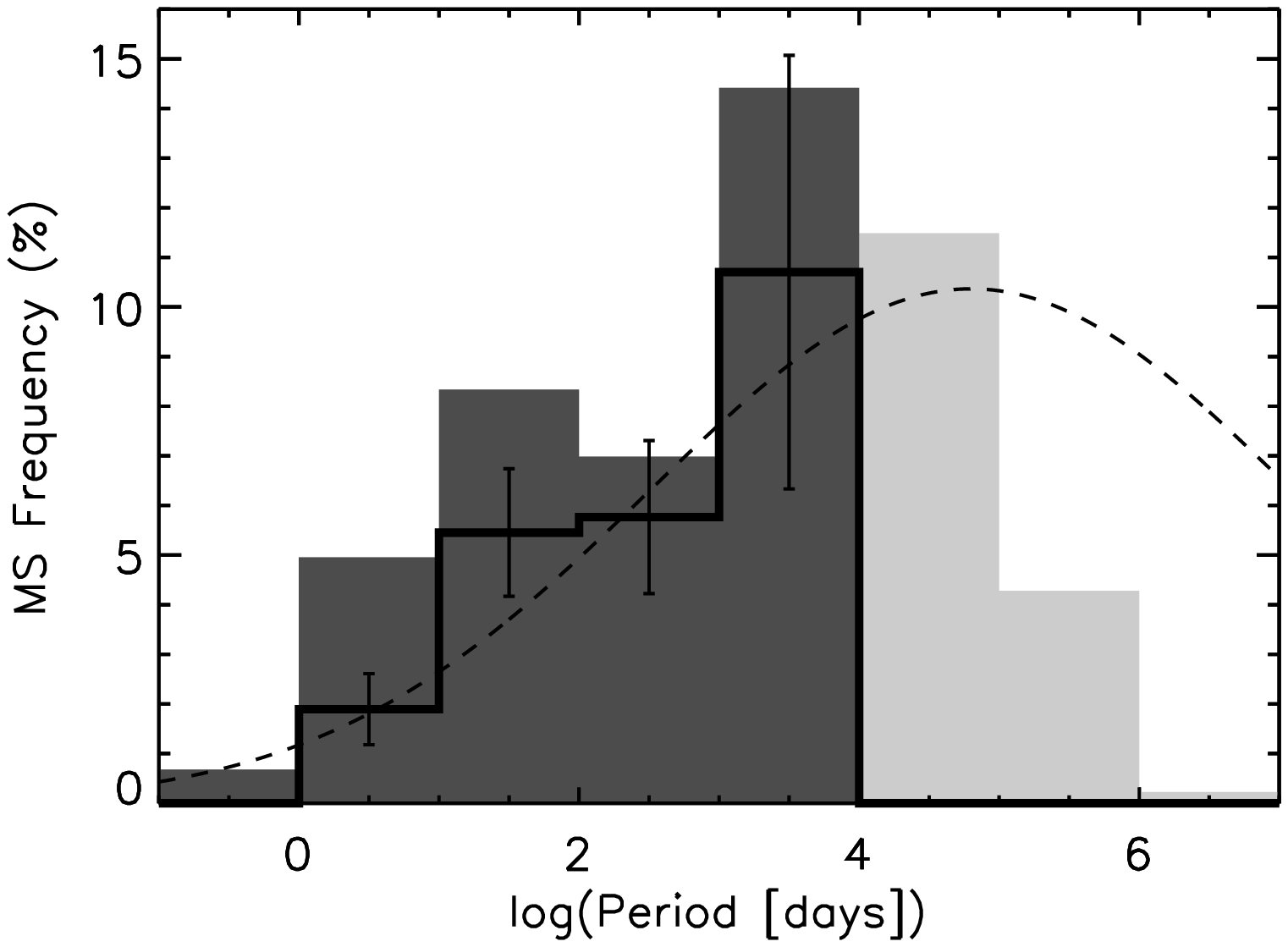} & \includegraphics[width=0.5\linewidth]{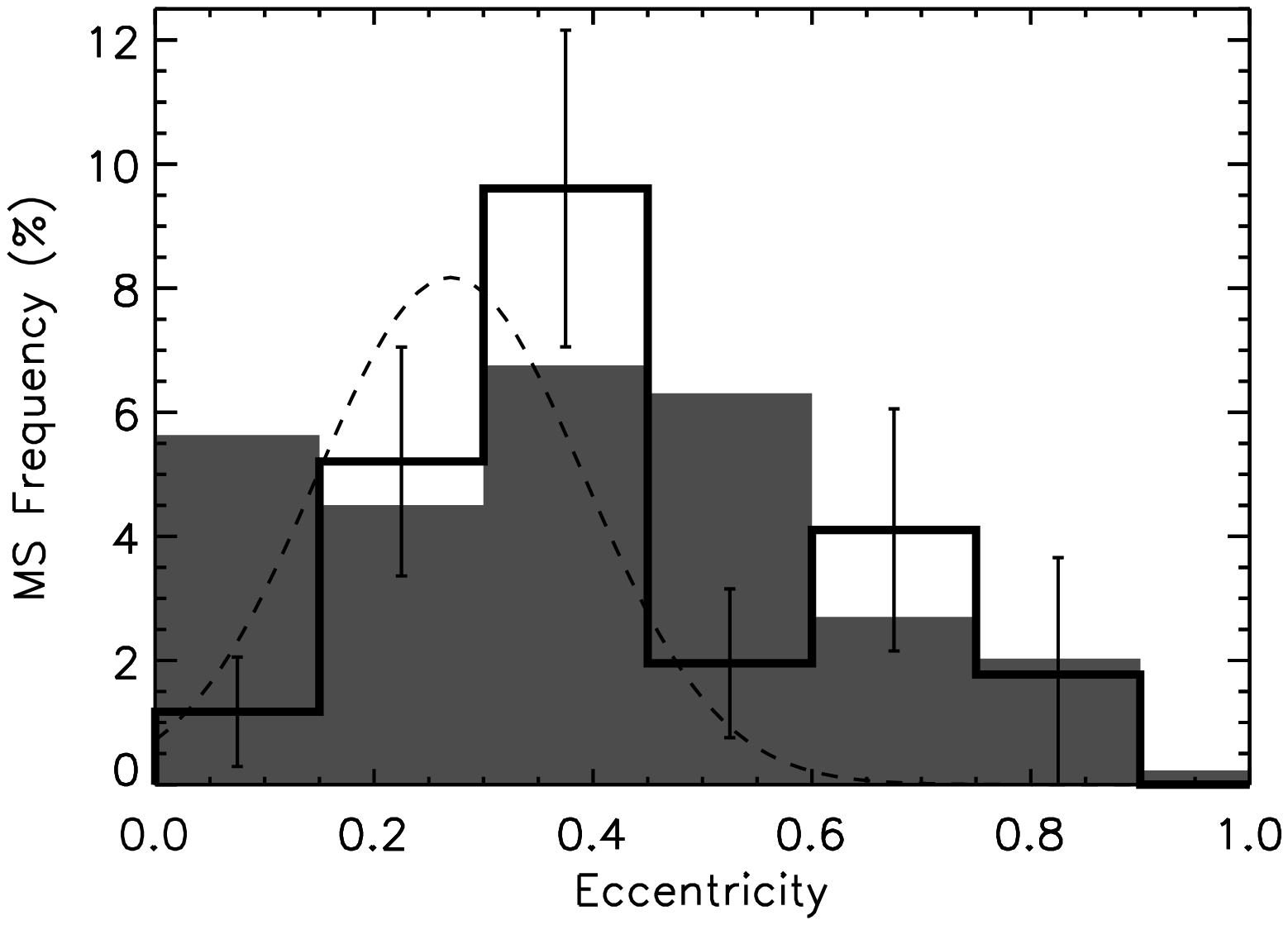} \\
\end{tabular}
\caption{\label{NGC188fig2}
Period (\textit{left}) and eccentricity (\textit{right}) MS binary frequency distributions at 7 Gyr.
The NGC 188 observed distributions are plotted in the solid black lines and have been corrected for 
incompleteness in the same manner as in Figure~\ref{M35fig}, showing all binaries with $P < 10^4$ days and 
excluding those with periods below the circularization period in the eccentricity distribution 
plot.  The simulated binary distributions are plotted in the filled histograms with the same gray-scale
coding as in Figure~\ref{M35fig}.  Again, we also show the \citet{duquennoy:91} Galactic field binary 
frequency distributions in the dashed lines for comparison.  Note the close agreement between 
the simulated and observed binaries in both distributions and binary frequency.}
\end{center}
\end{figure}

The WOCS observations of NGC 188 \citep{geller:08,geller:09a,geller:09b} allow for a detailed comparison 
with the NGC 188 model.  In the following analysis, we limit the output of our simulation to only include
stars within the Geller et al.~observable domain (in both mass and binary period).
We first focus on the color-magnitude diagram (CMD) at 7 Gyr shown in Figure~\ref{NGC188fig1}.
Note that the stellar evolution incorporated into the simulation does an excellent job of reproducing the 
``normal'' single and binary stars sequence as compared to the observed CMD.  However, the simulation
is notably lacking in BSs; there are 21 observed BSs in NGC 188 and only 5
at 7 Gyr in the simulation.  This low number of BSs stays relatively constant throughout the entire
simulation, and is likely due to the initial binary population.  We will discuss the implications of this 
result in Section~\ref{disc}.

We find the solar-type MS binary population in the simulation at 7 Gyr to correspond remarkably 
well to observations of similar binaries in NGC 188.
In Figure~\ref{NGC188fig2} we compare the observed incompleteness-corrected NGC 188 binaries (black lines) to 
those of the simulation at 7 Gyr (filled histograms) in both period and eccentricity.  The 
form of both the simulated distributions, as well as the 
binary frequency agree well with the observations of NGC 188 as well as the Galactic field \citep[dashed lines;][]{duquennoy:91}.
This close agreement is a testament to the power of using detailed observations to define the 
initial binary population in a $N$-body simulation.

Still, there are a few discrepancies between the simulated binary population and the observations.
The circularization period in the model is significantly shorter than the observed value for NGC 188
(Figure~\ref{NGC188fig3}).
\citet{meibom:05} find a circularization period 
of 14.5$\substack{+1.4 \\ -2.2}$~days, while a similar analysis of the model shows a period of 2.5 days. 
Interestingly, we also 
see \textit{too many} circular MS binaries in the simulation at 7 Gyr (first bin in Figure~\ref{NGC188fig2}), 
most having periods greater than the circularization period of both the simulation and observations.  
We discuss both of these results in the following section.

\section{Discussion and Future Directions}
\label{disc}

By utilizing the observed M35 binaries as a proxy for our primordial binary population,
we have managed to nearly reproduce the NGC 188 binaries within our $N$-body simulation at 7 Gyr.  
However, as noted in Section~\ref{comp188} the current NGC 188 model is 
deficient in BSs.  Interestingly, the \citet{hurley:05} model of M67 was able to produce 
the large number of observed BSs (though not their binary frequency), but was unable to 
reproduce the observed binary population.  The quantity of BSs produced in the M67 model 
was largely dependant on the initial binary population.  Hurley et al.~found that a large fraction 
of short-period binaries was necessary to create the observed numbers of BSs in M67 (and NGC 188).
These short-period binaries not only have a higher likelihood of creating BSs in isolation 
(e.g., through mass transfer resulting from Roche lobe overflow), but dynamical interactions involving 
short-period binaries have a higher probability of resulting in collisions and 
mergers \citep{fregeau:04}.
Observations of M35 binaries are very different from the flat period distribution and large binary 
frequency used in the M67 model (e.g., Figure~\ref{M35fig}).  
The important next step will be to match 
both the binary population and the BSs population within one cohesive simulation.  
We are currently exploring a number of possible methods for addressing this issue, 
the most promising of which appears to be the inclusion of primordial triples (e.g.,
Kozai induced mergers; \citealt{perets:09}, larger cross section for dynamical interactions, etc.).

\begin{figure}
\begin{center}
\includegraphics[width=0.8\linewidth]{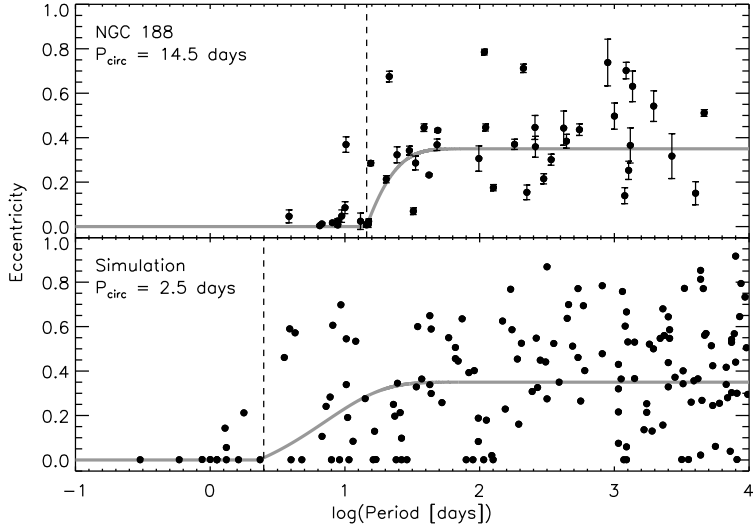}
\caption{\label{NGC188fig3}
Eccentricity plotted against the logarithm of the period ($e$ - log $P$) at 7 Gyr for the observed (top)
and simulated (bottom) solar-type MS NGC 188 binaries.  The solid
gray lines represent the best-fit circularization functions from \citet{meibom:05} for each sample, 
respectively.  The tidal processes in the simulation result in a circularization period that is 
significantly lower than in NGC 188.  Also, the simulation produces a population of circular binaries 
with $P >> P_{circ}$ that is not observed in NGC 188 (or the Galactic field).  Most of these
circular binaries in the model have previously gone through a stage of mass transfer.}
\end{center}
\end{figure}

The tidal circularization period is one of our best observational tools to study the effects of tides on 
a binary population.  In Figure~\ref{NGC188fig3} and Section~\ref{comp188} we show that the circularization
period at 7 Gyr in the simulation is significantly lower than what is observed in NGC 188.  This 
suggests that the tidal energy dissipation rate may be \textit{underestimated} in the $N$-body model 
(at least for solar-type MS binaries).  We may also (or alternatively) require a form of pre-MS tidal 
circularization \citep[e.g.][]{kroupa:95}.  As pointed out by \citet{mathieu:08}, tidal processes 
have a significant impact on dynamical interactions between single and binary stars, and the
energy dissipation rates can be the deciding factor between a close fly-by, tidal capture, or
even a merger or collision that could result in a BS.  WOCS data \citep[e.g.,][]{meibom:05}
provide the ideal foundation from which we will improve the tidal physics within the $N$-body model.

Finally, the simulation has an excess of circular binaries with periods beyond the circularization period.
The majority of these binaries have undergone some form of mass transfer, which has quickly 
circularized the orbit within the simulation.  
However, similar populations of circular binaries at $P >> P_{circ}$ are not observed
in open clusters or in the Galactic field \citep[e.g.,][]{meibom:05,duquennoy:91}.  This 
may suggest that mass transfer, especially in initially eccentric binaries, does not necessarily lead 
to circular binaries \citep[e.g.,][]{sepinsky:07,bonacic:08}, or that these cases of mass transfer happen 
much more frequently within the simulation than in reality.  Correctly modeling the products of 
mass transfer and common-envelope evolution is critical for accurately reproducing BSs 
as well as the binary population within an $N$-body simulation.

We will present further details about this NGC 188 model in future papers.  Additionally, we will explore the 
various methods mentioned above for bringing the NGC 188 model, and $N$-body simulations in general, in better
agreement with observations.  This synergy between observations and simulations will continue to refine the
$N$-body method and reveal further insights into the origins of BSs, the evolution of 
a binary population, and the dynamical evolution of star clusters.

\vspace{1em}
\footnotesize \flushleft
This work was funded in part by a National Science Foundation (NSF) East Asia and Pacific Summer Institutes (EAPSI)
fellowship, the Wisconsin Space Grant Consortium, and NSF grant AST-0406615.


\end{document}